\documentclass[twocolumn,english,prl,twocolumn]{revtex4}
\usepackage[T1]{fontenc}
\usepackage[latin1]{inputenc}
\usepackage{graphicx}

\makeatletter

\usepackage{babel}
\makeatother
\begin{document}

\title{Is $\Delta_{\pi}$-gap-only superconductivity possible in $Mg_{1-x}Al_{x}B_{2}$
and $Mg(B_{1-y}C_{y})_{2}$ alloys ?}

\author{P. Jiji Thomas Joseph and Prabhakar P. Singh}

\affiliation{Department of Physics, Indian Institute of Technology Bombay, Mumbai
- 400076, India}

\begin{abstract}
Using density-functional-based method, we study the $\mathbf{k}$-resolved
$\sigma$- and $\pi$-band holes in $Mg_{1-x}Al_{x}B_{2}$ and $Mg(B_{1-y}C_{y})_{2}$
alloys. We find that the calculated profiles of the loss of $\sigma$-
and $\pi$-band holes in these two systems as a function of impurity
concentration are in qualitative agreement with experiments, as expected.
We also describe its implications \emph{vis-a-vis} superconductivity
in $Mg_{1-x}Al_{x}B_{2}$ and $Mg(B_{1-y}C_{y})_{2}$.
\end{abstract}
\maketitle
The surprising discovery of superconductivity at $40\, K$ in a simple
metal $MgB_{2}$ \cite{nature-410-63} a few years ago continues to
test the details of the microscopic picture offered by Bardeen-Cooper-Schriefer
theory of superconductivity and its extensions thereof. Generally,
it is agreed that in $MgB_{2}$ (i) there are two superconducting
gaps \cite{nature-423-65,prl-89-187002,prl-89-257001,prl-91-127001,prl-87-137005,prl-88-127002,prl-87-087005}
$\Delta_{\sigma}$ and $\Delta_{\pi}$ of 7 meV and 2 meV, respectively,
(ii) the two gaps arise from the holes in the $B$ $p_{\sigma}$ and
$p_{\pi}$ bands \cite{prl-87-087005,prb-63-220504,prl-86-4371,nature-418-758,prb-65-132518,pr-175-537,prb-65-012505},
(iii) the holes in the $B$ $p_{\sigma}$ and $p_{\pi}$ bands couple
strongly to the in-plane $E_{2g}$ phonon mode \cite{prb-65-132518,prl-86-4656,prl-86-5771,prb-66-020513},
and (iv) the coupling is particularly strong along $\Gamma$-$A$
direction and around $M$ point in the hexagonal Brillouin zone \cite{pps_ep_gtoa}. 

To gain further insight into the role played by the holes of the $B$
$p_{\sigma}$ and $p_{\pi}$ bands in deciding the superconducting
properties as well as the relative strengths of the two gaps $\Delta_{\sigma}$
and $\Delta_{\pi}$ in $MgB_{2}$, recently, there has been a concerted
effort to examine the various superconducting properties of \emph{doped}
$MgB_{2}$. In particular, the $Al$ doping at the $Mg$ site and
the $C$ doping at the $B$ site in $MgB_{2}$ \cite{nature-410-343,jpcm-13-11689,prb-65-214536,prb-65-132505,prb-65-020507,prb-66-014518,prb-66-140514,prb-65-174515,physC-384-227,physC-385-16,prb-68-060508}
are expected to fill up the holes in the $B$ $p_{\sigma}$ and $p_{\pi}$
bands which, in turn, would affect the two gaps $\Delta_{\sigma}$
and $\Delta_{\pi}$. Understanding the changes in the superconducting
properties of $Mg_{1-x}Al_{x}B_{2}$ as a function of $Al$ concentration
$x$ and of $Mg(B_{1-y}C_{y})_{2}$ as a function of $C$ concentration
$y$ would provide a more detailed information about the electron-phonon
and the electron-electron interactions in these alloys. 

Experimentally, in $Mg_{1-x}Al_{x}B_{2}$ the superconducting transition
temperature $T_{c}$ as well as the gap $\Delta_{\sigma}$ are found
to decrease with increasing $Al$ concentration \cite{nature-410-343,prb-65-132505,prb-71-144505,prb-71-174506,prb-65-214536}.
The change in the gap $\Delta_{\pi}$ is relatively small up to $x\,~\,0.7$
at which the superconductivity vanishes in $Mg_{1-x}Al_{x}B_{2}$.
However, in $Mg(B_{1-y}C_{y})_{2}$ the $T_{c}$ \cite{cmat-0405060,physC-355-1,physC-385-16,prb-64-134513,prb-65-052505,prb-70-024504,prb-70-064520,prb-71-020501,prb-71-134511,prl-92-217003}
and the $\Delta_{\sigma}$ decrease rapidly as a function of $C$
concentration with superconductivity vanishing at around $y\,~\,0.15$.
The change in $\Delta_{\pi}$, once again, is found to be minimal
in $Mg(B_{1-y}C_{y})_{2}$ \cite{prb-68-020505,prb-68-060508,prb-70-064520,prb-71-024533,prb-71-134511}.

Most of the previous theoretical studies \cite{prb-66-012511,physC-382-381,ssc-127-271,prb-67-132509,prb-68-144508,physC-407-121,cmat-0409563,prl-94-027002,prl-95-267002}
of $Mg_{1-x}Al_{x}B_{2}$ and $Mg(B_{1-y}C_{y})_{2}$ used virtual-crystal
approximation \cite{prb-66-012511,prb-68-144508,physC-407-121,prl-94-027002,prl-95-267002}
or supercell approach \cite{prb-67-132509}, having limited predictive
capability, to describe the chemical disorder in the $Mg$ sub-lattice
and the $B$ sub-lattice respectively. An accurate and reliable description
of chemical disorder, provided by coherent-potential approximation,
has been used previously in the study of $Mg_{1-x}Al_{x}B_{2}$ and
$Mg(B_{1-y}C_{y})_{2}$ \cite{physC-382-381,ssc-127-271,cmat-0409563}
but these studies provided either $\mathbf{k}$-integrated information
or were inconclusive. 

Since many of the superconducting properties of $MgB_{2}$ and its
alloys $Mg_{1-x}Al_{x}B_{2}$ and $Mg(B_{1-y}C_{y})_{2}$ are crucially
dependent on their \emph{normal state} electronic structure \cite{prl-86-4656,prl-86-5771,prl-87-037001,prl-87-087005,prl-89-107002,prl-87-087004},
especially along $\Gamma$-$A$ direction and $M$ point of the hexagonal
Brillouin zone (BZ), we study the changes in $Mg_{1-x}Al_{x}B_{2}$
and $Mg(B_{1-y}C_{y})_{2}$ alloys as a function of $Al$ and $C$
concentrations respectively, using $\mathbf{k}$-resolved Bloch spectral
function $A(\mathbf{k},E)$ \cite{prb-21-3222,ProgMater-27-1} and
$B$ and $C$ $p_{x(y)}$ and $p_{z}$ densities of states. It is
expected that with increasing $Al$ and $C$ concentrations the holes
in the respective alloys would get gradually filled up. However, not
surprisingly, the way the holes are filled up in $Mg_{1-x}Al_{x}B_{2}$
and $Mg(B_{1-y}C_{y})_{2}$ alloys make the two systems quite different
from each other as detailed below.

In this Letter we show that in $Mg_{1-x}Al_{x}B_{2}$ (i) the $\sigma$-band
holes along $\Gamma$-$A$ get annihilated by $x\,~\,0.8$ with the
holes at $\Gamma$ vanishing first at $x\,~\,0.5$, and (ii) the $\pi$
holes at $M$ vanish by $x\,~0.7$. However, in $Mg(B_{1-y}C_{y})_{2}$
(i) the $\sigma$-band holes along $\Gamma$-$A$ deplete rapidly
and vanish beyond $y\,\sim\,0.25$ with the holes at $A$ vanishing
first, and (ii) the $\pi$-band holes at $M$ as well as $\sigma$-band
holes at $A$ reduce substantially beyond $y\,\sim\,0.1.$ Before
we discuss our results, we briefly describe the details of our calculations. 

The normal metal electronic structure of the disordered alloys $Mg_{1-x}Al_{x}B_{2}$
and $Mg(B_{1-y}C_{y})_{2}$ are calculated using the Korringa-Kohn-Rostoker
(KKR) Green's function method formulated in the atomic sphere approximation
(ASA) \cite{turek,phariseau}, which has been corrected by the use
of both the muffin-tin correction for the Madelung energy, needed
for obtaining an accurate description of ground state properties in
the ASA \cite{prl-55-600}, and the multipole moment correction to
the Madelung potential and energy which significantly improves the
accuracy by taking into consideration the non-spherical part of polarization
effects \cite{cmc-15-119}. Chemical disorder was taken into account
by means of coherent-potential approximation (CPA) \cite{pr-156-809}.
The exchange and correlation were included within the local density
approximation (LDA) using the Perdew-Wang parameterization of the
many-body calculations of Ceperley and Alder \cite{prb-45-13244}.
During the self-consistent procedure the reciprocal space integrals
were calculated by means of $2299$ $\mathbf{k}$-points in the irreducible
part of the hexagonal BZ, while the energy integrals were evaluated
on a semicircular contour in the complex energy plane using $24$
energy points distributed in such a way that the sampling near the
Fermi energy was increased. When calculating the density of states
the number of $\mathbf{k}$-points were increased to $6075$. The
atomic sphere radii of $Mg$ and $B$ were kept as $1.294$ and $0.747$,
respectively of the Wigner-Seitz radius. The sphere radii of the substituted
$Al$ and $C$ were kept the same as that of $Mg$ and $B$ respectively.
The overlap volume resulting from the blow up of the muffin-tin spheres
was approximately $10$\%, which is reasonable within the accuracy
of the approximation \cite{skriver-1984}. The charge self-consistency
iterations were accelerated using the Broyden's mixing scheme \cite{prb-38-12807},
the CPA self consistency was accelerated using the prescription of
Abrikosov $et$ $al$ \cite{PRB-56-9319}. The calculations for $Mg_{1-x}Al_{x}B_{2}$
for $0\leq x\leq1$ and $Mg(B_{1-y}C_{y})_{2}$ for $0\leq y\leq0.3$
were carried out at experimentally obtained lattice parameters of
these alloys , assuming a rigid underlying lattice. The effects of
local lattice-relaxation as well as any possible short-range ordering
effects are not considered. Our results for $Mg_{1-x}Al_{x}B_{2}$
and $Mg(B_{1-y}C_{y})_{2}$ are analyzed in terms of Bloch spectral
density, $A(\mathbf{k},E)$ \cite{prb-21-3222} and the $B$ and $C$
$p$ densities of states.

In general, electron doping of $MgB_{2}$ by substituting $Al$ at
the $Mg$ sub-lattice or $C$ at the $B$ sub-lattice can have three
important effects, (i) the outward movement of the Fermi energy due
to addition of electrons, (ii) disorder-induced broadening of peaks
in $A(\mathbf{k},E)$, and (iii) redistribution of states due to hybridization.
If the disorder and hybridization-related effects are negligible then
electron doping can be well described by a rigid-band-like picture,
where $E_{F}$ is shifted up using $MgB_{2}$ densities of states
until the added electrons are accommodated. In such a case, the $\sigma$-
and $\pi$-band holes would gradually get filled up as a function
of electron doping. Thus, in the rigid-band picture the $\sigma$-
and $\pi$-band-hole-dependent superconducting properties of $MgB_{2}$
alloys would be similar as a function of electron doping. However,
if the disorder and the hybridization lead to significant modifications
of the $\sigma$- and $\pi$-band holes then the electronic properties
of both the normal and the superconducting states of $MgB_{2}$ alloys
would depend on the details of electron doping rather than just the
electron count. Experimentally, as outlined in earlier paragraphs,
the $\sigma$- and $\pi$-band-hole-dependent properties of the two
alloys, $Mg_{1-x}Al_{x}B_{2}$ and $Mg(B_{1-y}C_{y})_{2}$, are found
to differ significantly from each other as a function of $Al$ and
$C$ concentrations, underscoring the importance of disorder and hybridization
effects in these alloys, in particular on $\sigma$- and $\pi$-band
holes. Therefore, in the following, we primarily focus on the changes
in the states close to $E_{F}$, which happen to be the $p_{\sigma}$
states at $\Gamma$ and $A$ points and $p_{\pi}$ state at $M$ point
of the hexagonal BZ. 

\begin{figure}[h]
\includegraphics[%
  clip,
  scale=0.33]{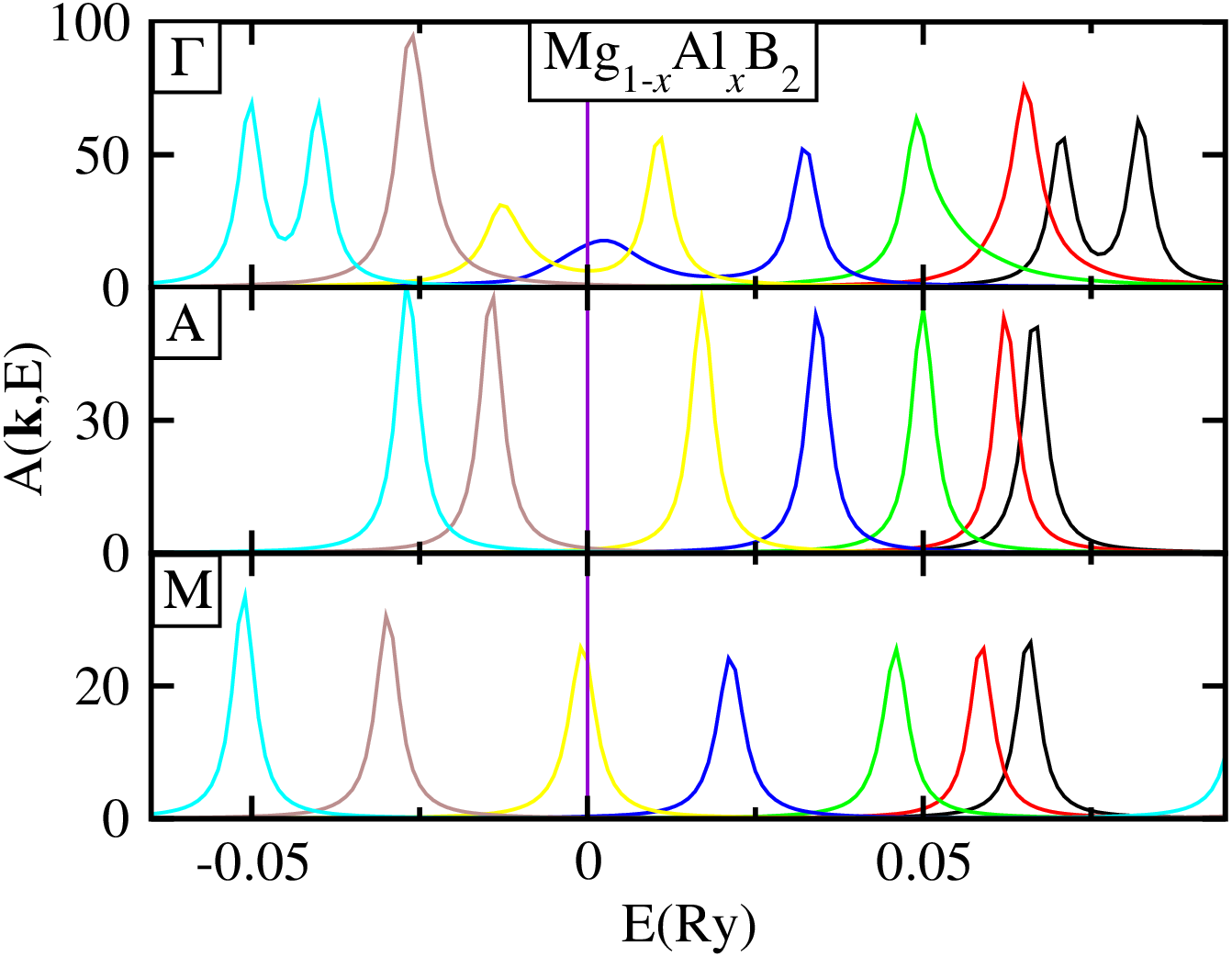}

\caption{\label{spec-al}The Bloch spectral function $A(\mathbf{k},E)$ of
$Mg_{1-x}Al_{x}B_{2}$ alloys calculated at $\Gamma$ (top panel),
$A$ (middle panel) and $M$ (bottom panel) points of the hexagonal
Brillouin zone using the KKR-ASA-CPA method for $x$ equal to $0$
(black), $0.1$ (red), $0.3$ (green), $0.5$ (blue), $0.7$ (yellow),
$0.9$ (brown) and $1$ (cyan), respectively. The vertical line through
the energy zero represents the Fermi energy. }
\end{figure}

The most important result of the present work for $Mg_{1-x}Al_{x}B_{2}$
alloys is shown in Fig.\ref{spec-al}, where we show the Bloch spectral
function $A(\mathbf{k},E)$ as a function of $x$ for $0\,\leq\, x\,\leq\,1$
at $\Gamma$, $A$ and $M$ points of the hexagonal BZ in a limited
energy window ($\pm0.1$ Ry) around Fermi energy $E_{F}$. Along $\Gamma$-$A$,
the doubly degenerate $B$ $p_{\sigma}$ states in $Mg_{1-x}Al_{x}B_{2}$
alloys move closer to $E_{F}$ with increasing $Al$ concentration
from $x=0.1$ to $x=0.70$, indicating a decrease in $\sigma$-band
holes and subsequently, weakening of electron-phonon coupling along
this direction. We also find that for $x\sim0.5$, the peak in $A(\mathbf{k},E)$
is at $E_{F}$ indicating some topological changes in the Fermi surface
of the alloy. By $x=0.90$, the $B$ $p_{\sigma}$ states at $A$
are well inside $E_{F}$, resulting in the annihilation of all the
$p_{\sigma}$ hole states along $\Gamma$-$A$. Thus, we should not
expect $\sigma$-band-hole-based superconductivity beyond $x=0.80$.
We also find that the $Mg$-derived antibonding $s$-state ($\Gamma_{6})$
moves closer to the $p_{\sigma}$ state for intermediate concentrations
but separates out for $x=1.0$. 

The $B$ $p_{\pi}$ hole states at $M$ in $Mg_{1-x}Al_{x}B_{2}$
alloys get gradually annihilated with increasing $Al$ concentration
from $x=0.1$ to $x=0.50$. We find that the $A(\mathbf{k},E)$ peak
corresponding to $p_{\pi}$ states crosses $E_{F}$ at a lower $x$
than the $p_{\sigma}$ states at $A$ point. As the contribution to
the electron-phonon coupling responsible for $\pi$-band superconductivity
comes from $M$ and other similar points in the BZ, a $\Delta_{\pi}$-gap-only
superconductivity seems unlikely in $Mg_{1-x}Al_{x}B_{2}$ alloys.
However, it does not preclude a crossover from a $\sigma$ to $\pi$
dominated superconductivity \cite{prl-95-267002}. 

\begin{figure}[h]
\includegraphics[%
  clip,
  scale=0.33]{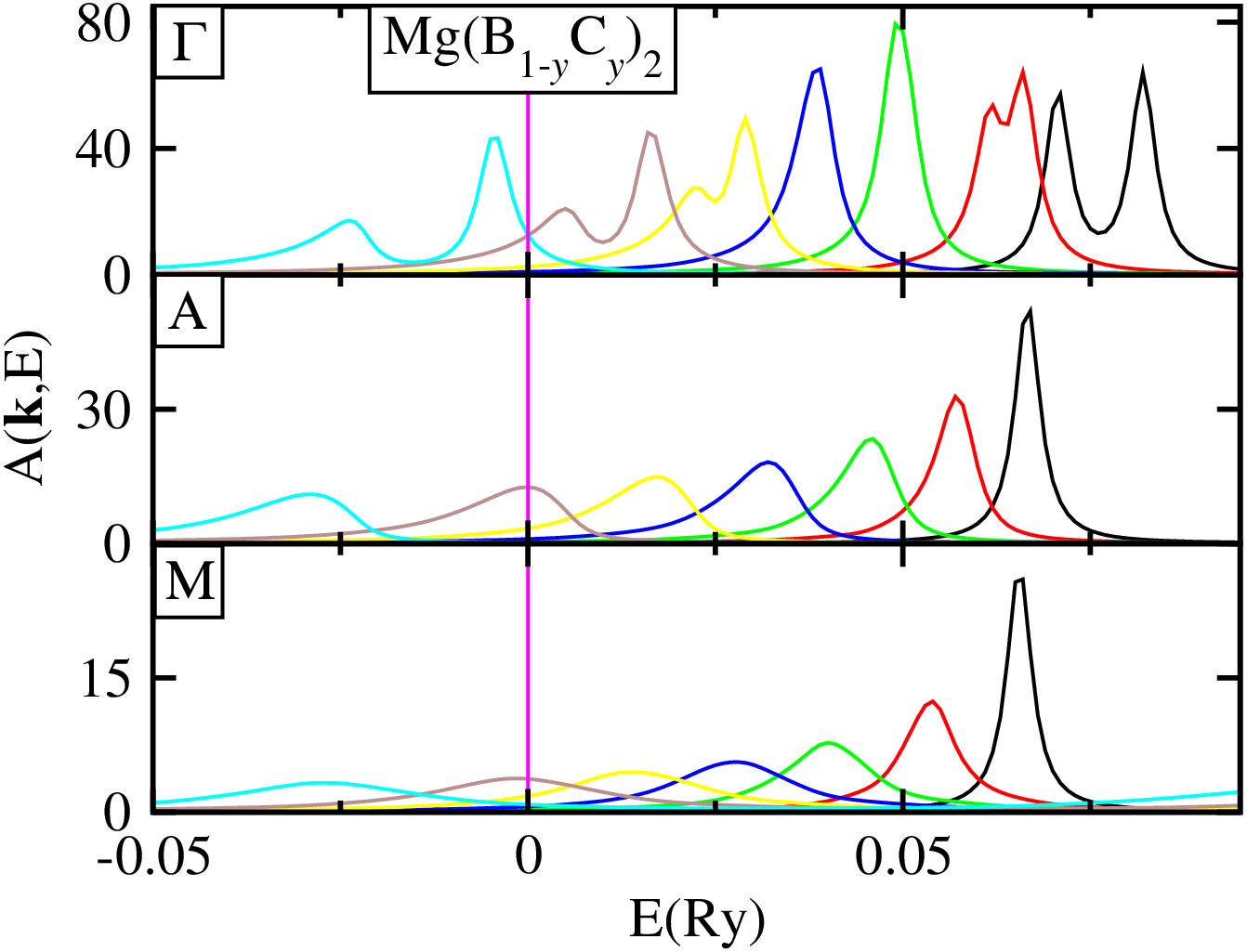}

\caption{\label{spec-c}The Bloch spectral function $A(\mathbf{k},E)$ of
$Mg(B_{1-y}C_{y})_{2}$ alloys calculated at $\Gamma$ (top panel),
$A$ (middle panel) and $M$ (bottom panel) points of the hexagonal
Brillouin zone using the KKR-ASA-CPA method for $y$ equal to $0$
(black), $0.05$ (red), $0.1$ (green), $0.15$ (blue), $0.2$ (yellow),
$0.25$ (brown) and $0.3$ (cyan), respectively. The vertical line
through the energy zero represents the Fermi energy. }
\end{figure}

In Fig. \ref{spec-c} we show the Bloch spectral function $A(\mathbf{k},E)$
of $Mg(B_{1-y}C_{y})_{2}$ alloys as a function of $y$ for $0\,\leq\, y\,\leq\,0.3$
at $\Gamma$, $A$ and $M$ points around Fermi energy. We find that
with increasing $C$ concentration from $y=0.05$ to $y=0.2$ the
$p_{\sigma}$ states along $\Gamma$-$A$ move towards $E_{F}$ as
well as get redistributed on the energy scale due to disorder and
hybridization. In fact, in the range $y=0.1$ to $y=0.15$, we find
substantial reduction in $p_{\sigma}$ states around $A$ point. Since
the electron-phonon coupling is stronger near $A$ point, the loss
of $p_{\sigma}$ states should dramatically affect the $\sigma$-band
superconductivity. Similar to the reduction of $p_{\sigma}$ states
along $\Gamma$-$A$, the $p_{\pi}$ hole states at $M$ point in
$Mg(B_{1-y}C_{y})_{2}$ alloys are drastically reduced by $y=0.1$.
Here, it must be noted that the reduction in $\sigma$- and $\pi$-band
holes in $Mg_{1-x}Al_{x}B_{2}$ with increasing $Al$ concentration
is mostly due to the upward movement of the Fermi energy. In contrast,
disorder and hybridization play a dominant role in reducing the $\sigma$-
and $\pi$-band holes in $Mg(B_{1-y}C_{y})_{2}$ alloys. 

\begin{figure}
\includegraphics[%
  clip,
  scale=0.33]{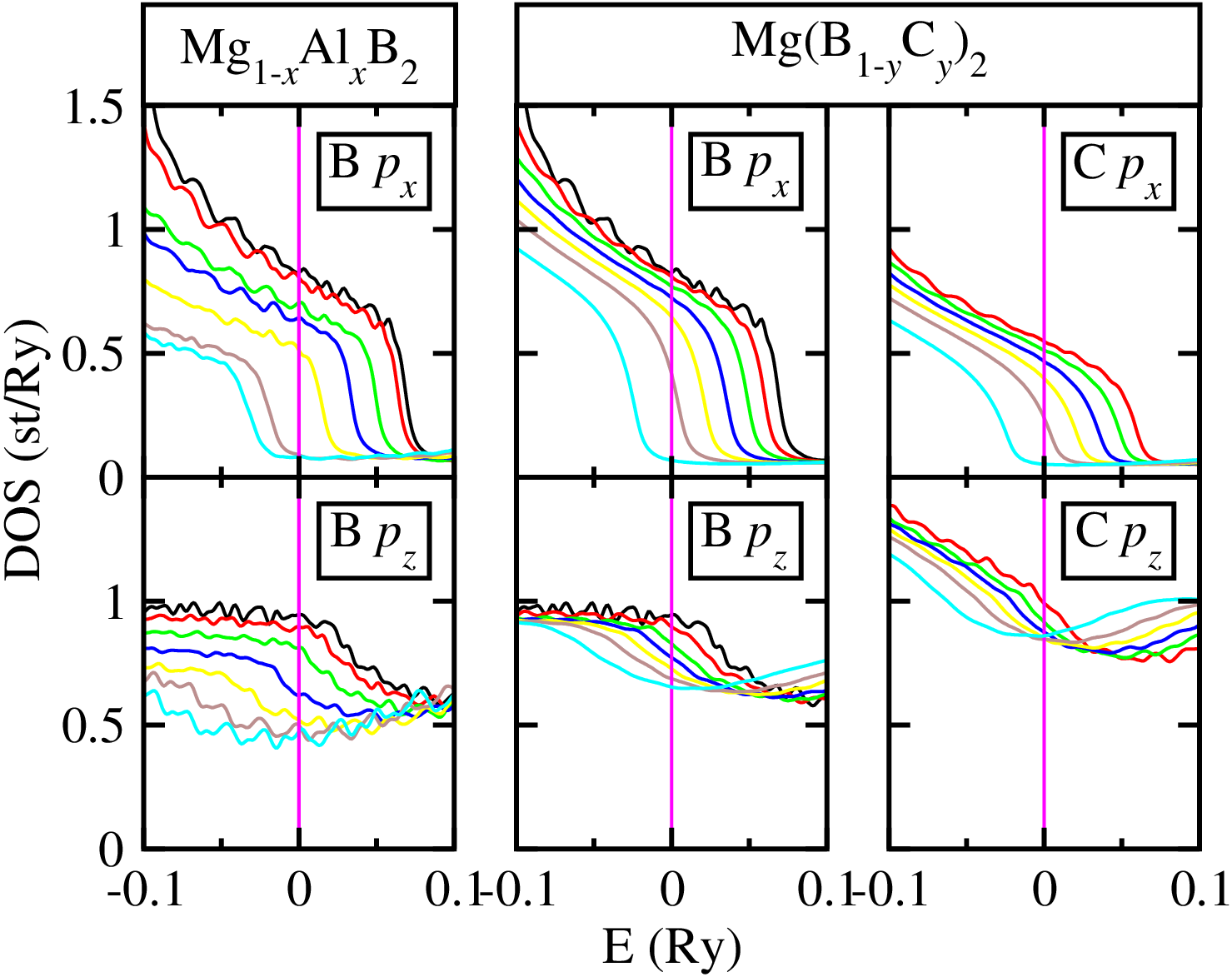}

\caption{\label{dos-compa}The atom ($B$ and $C$) and orbital ($p_{x(y)}$and
$p_{z}$) projected density of states (DOS) in $Mg_{1-x}Al_{x}B_{2}$
(left panels) and $Mg(B_{1-y}C_{y})_{2}$ (middle and right panels)
alloys calculated using the KKR-ASA-CPA method for $x(y)$ equal to
$0(0)$ (black), $0.1(0.05)$ (red), $0.3(0.1)$ (green), $0.5(0.15)$
(blue), $0.7(0.2)$ (yellow), $0.9(0.25)$ (brown) and $1(0.3)$ (cyan),
respectively. The vertical line through the energy zero represents
the Fermi energy. }
\end{figure}

The changes in the $B$ $p_{\sigma}$ and $p_{\pi}$ hole states in
$Mg_{1-x}Al_{x}B_{2}$ as a function of $Al$ concentration and in
$Mg(B_{1-y}C_{y})_{2}$ as a function of $C$ concentration can be
seen more clearly in their respective partial densities of states
as shown in Fig. \ref{dos-compa}. In $Mg_{1-x}Al_{x}B_{2}$ alloys
the $B$ $p_{x(y)}$ hole states decrease steadily as a function of
$x$, vanishing for $x\sim0.8$ as shown in Fig. \ref{dos-compa}.
Thus, as stated earlier in the context of $A(\mathbf{k},E)$, we should
not expect $\sigma$-band-hole-based superconductivity beyond $x=0.80$
$Mg_{1-x}Al_{x}B_{2}$. However, the decrease in $B$ $p_{z}$ hole
states with increasing $x$ is not as much. Even for $x=1.0$, there
are $B$ $p_{z}$ hole states available in $Mg_{1-x}Al_{x}B_{2}$
as shown in Fig.\ref{dos-compa} but it does not sustain $\Delta_{\pi}$-gap-only
superconductivity. 

For $Mg(B_{1-y}C_{y})_{2}$ alloys, our results show that $p_{x(y)}$
and $p_{z}$ densities of states of both $B$ and $C$ behave similarly
with increasing $C$ concentration, as can be seen from Fig. \ref{dos-compa}.
By $y\sim0.25$, most of the $p_{x(y)}$ hole states of $B$ and $C$
have moved inside $E_{F}$. As pointed out above, in $Mg(B_{1-y}C_{y})_{2}$
alloys with the addition of $C$ those $\sigma$-band holes are lost
first which couple to the phonons more strongly which may lead to
a loss of superconductivity in spite of having some $p_{x(y)}$ hole
states of $B$ and $C$. The changes in the $p_{z}$ hole states are
minimal in $Mg(B_{1-y}C_{y})_{2}$, with a slight increase in the
hole states for higher $C$ concentrations. 

We like to point out that the approximations used in the present work,
namely the use of spherically symmetric potential to describe the
effective one-electron interaction and the overlapping atomic spheres
to fill the space, are expected to affect some numerical aspects of
the present work. The use of optimized lattice constants is expected
to similarly affect the present work, especially in $Mg(B_{1-y}C_{y})_{2}$
alloys for $y\geq0.2$ because for these concentrations we have used
the lattice constants corresponding to $y=0.15$. However, our results
for $MgB_{2}$ and $AlB_{2}$ are consistent with the more accurate,
full-potential results, and since the coherent-potential approximation
is known to describe the effects of disorder reliably, we expect our
intermediate-concentration-results to be robust and qualitatively
correct.

In conclusion, we have studied the $\mathbf{k}$-resolved $\sigma$-
and $\pi$-band holes in $Mg_{1-x}Al_{x}B_{2}$ and $Mg(B_{1-y}C_{y})_{2}$
alloys. The calculated profiles of the loss of $\sigma$- and $\pi$-band
holes in these two systems as a function of impurity concentration
are expected to be in qualitative agreement with the experiments.
We have also described its implications \emph{vis-a-vis} superconductivity
in $Mg_{1-x}Al_{x}B_{2}$ and $Mg(B_{1-y}C_{y})_{2}$ alloys.

One of us (PJTJ) would like to thank Dr. Andrei Ruban for providing
the KKR-ASA code. Discussions with Dr. Igor Mazin on the electronic
structure properties of $MgB_{2}$ is gratefully acknowledged.


\begin{thebibliography}{1}
\bibitem{nature-410-63}J. Nagamatsu \emph{et al}., Nature 410, 63 ( 2001)
\bibitem{nature-423-65}S. Souma \emph{et al}., Nature 423, 65 (2003)
\bibitem{prl-89-187002}M. Iavarone \emph{et al}., Phys. Rev. Lett. 89, 187002 (2002) 
\bibitem{prl-89-257001}F. Bouquet \emph{et al}., Phys. Rev. Lett. 89, 257001 (2002) 
\bibitem{prl-91-127001}S. Tsuda \emph{et al}., Phys. Rev. Lett. 91, 127001 (2003) 
\bibitem{prl-87-137005}P. Szabo \emph{et al}., Phys. Rev. Lett. 87, 137005 (2001)
\bibitem{prl-88-127002}H. Schmidt \emph{et al}., Phys. Rev. Lett. 88, 127002 (2002) 
\bibitem{prl-87-087005}A. Y. Liu \emph{et al}., Phys. Rev. Lett. 87, 087005 (2001)
\bibitem{prb-63-220504}H. Schmidt \emph{et al}., Phys. Rev. B 63, 220504 (2001)
\bibitem{prl-86-4371}G. Karapetrov \emph{et al}., Phys. Rev. Lett. 86, 4374 (2001)
\bibitem{nature-418-758}H. J. Choi, D. Roundy \emph{et al}., Nature 418, 758 (2002)
\bibitem{pr-175-537}W. L. McMillan, Phys. Rev. 175, 537 (1968)
\bibitem{prb-65-012505}P. Seneor \emph{et al}., Phys. Rev. B 65, 012505 (2001) 
\bibitem{prb-65-132518}C. Joas \emph{et al}., Phys. Rev. B 65, 132518 (2002) 
\bibitem{prl-86-4656}J. Kortus \emph{et al}., Phys. Rev. Lett. 86, 4656 (2001)
\bibitem{prb-66-020513}H. J. Choi, D. Roundy \emph{et al}., Phys. Rev. B 66, 020513 (2002)
\bibitem{prl-86-5771}K.-P. Bohnen \emph{et al}., Phys. Rev. Lett. 86, 5771-5774 (2001)
\bibitem{pps_ep_gtoa}P. P. Singh, Phys. Rev. B 67, 132511 (2003)
\bibitem{nature-410-343}J. S. Slusky \emph{et al}., Nature 410, 343 (2001)
\bibitem{jpcm-13-11689}S. Agrestini \emph{et al}., J. Phys.: Condens. Matter 13, 11689 (2001)
\bibitem{prb-65-214536}J. Y. Xiang \emph{et al}., Phys. Rev. B 65, 214536 (2002)
\bibitem{prb-65-132505}J. Q. Li \emph{et al}., Phys. Rev. B 65, 132505 (2002)
\bibitem{prb-65-020507}P. Postorino \emph{et al}., Phys. Rev. B 65, 020507 (2001)
\bibitem{prb-66-014518}S. Margadonna \emph{et al}., Phys. Rev, B 66, 014518 (2002)
\bibitem{prb-66-140514}G. Papavassiliou \emph{et al}., Phys. Rev, B 66, 140514 (2002)
\bibitem{prb-65-174515}A. Bianconi \emph{et al}., Phys. Rev. B 65, 174515 (2002)
\bibitem{physC-385-16}R. A. Ribeiro \emph{et al}., Physica C 385, 16 (2003)
\bibitem{prb-68-060508}H. Schmidt \emph{et al}., Phys. Rev. B 68, 060508 (2003) 
\bibitem{physC-384-227}R. A. Ribeiro \emph{et al}., Physica C 384, 227 (2003) 
\bibitem{prb-71-174506}J. Karpinski \emph{et al}., Phys. Rev. B 71, 174506 (2005)
\bibitem{prb-71-144505}M. Putti \emph{et al}., Phys. Rev. B 71, 144505 (2005) 
\bibitem{physC-355-1}M Paranthaman \emph{et al}., Physica C 355, 1 (2001)
\bibitem{prb-65-052505}W. Mickelson \emph{et al}., Phys. Rev. B 65, 052505 (2002). 
\bibitem{cmat-0405060}S. M. Kazakov \emph{et al}., cond-mat/0405060
\bibitem{prb-64-134513}T. Takenobu \emph{et al}., Phys. Rev. B 64, 134513 (2001) 
\bibitem{prb-70-024504}T. Masui \emph{et al}., Phys. Rev. B 70, 024504 (2004)
\bibitem{prl-92-217003}R. H. T. Wilke \emph{et al}., Phys. Rev. Lett. 92, 217003 (2004)
\bibitem{prb-71-134511}G. A. Ummarino \emph{et al}., Phys. Rev. B 71, 134511 (2005) 
\bibitem{prb-71-020501}A. V. Sologubenko \emph{et al}., Phys. Rev. B 71, 020501 (2005)
\bibitem{prb-70-064520}Z. Holanova \emph{et al}., Phys. Rev. B 70, 064520 (2004)
\bibitem{prb-71-024533}S. M. Kazakov \emph{et al}., Phys. Rev. B 71, 024533 (2005)
\bibitem{prb-68-020505}P. Samuely \emph{et al}., Phys. Rev. B 68, 020505 (2003)
\bibitem{prb-66-012511}O. de la Pena \emph{et al}.,Phys. Rev. B 66, 012511 (2002)
\bibitem{physC-382-381}P. P. Singh, Physica C 382, 381 (2002)
\bibitem{ssc-127-271}P. P. Singh, Solid State Commun. 127, 271 (2003)
\bibitem{prb-67-132509}A. Bussmann-Holder and A. Bianconi, Phys. Rev. B 67, 132509 (2003)
\bibitem{prb-68-144508}G. Profeta \emph{et al}., Phys. Rev. B 68, 144508 (2003)
\bibitem{physC-407-121}G. A. Ummarino \emph{et al}., Physica C 407, 121 (2004)
\bibitem{cmat-0409563}D. Kasinathan \emph{et al}., cond-mat/0409563
\bibitem{prl-94-027002}J. Kortus \emph{et al}., Phys. Rev. Lett. 94, 027002 (2005)
\bibitem{prl-95-267002}L. D. Cooley \emph{et al}., Phys. Rev. Lett. 95, 267002 (2005)
\bibitem{prl-89-107002}I. I. Mazin \emph{et al}., Phys. Rev. Lett. 89, 107002 (2002)
\bibitem{prl-87-037001}T. Yildirim \emph{et al}., Phys. Rev. Lett. 87, 037001 (2001)
\bibitem{prl-87-087004}P. P. Singh, Phys. Rev. Lett. 87, 087004 (2001)
\bibitem{prb-21-3222}J. S. Faulkner and G. M. Stocks Phys. Rev. B 21, 3222 (1980)
\bibitem{ProgMater-27-1}J. S. Faulkner, Prog. Mater. Sci. 27, 1 (1982)
\bibitem{turek}I. Turek \emph{et al}., Electronic structure of disordered alloys,
surfaces and interfaces, (Kluwer Academic Publishers, Boston, 1997)
\bibitem{phariseau}Electrons in disordered metals and at metallic surfaces, P. Phariseau,
B. L. Gyorffy and L. Scheire, eds. New York, Plenum Press  (1979)
\bibitem{prl-55-600}N. E. Christensen and S. Satpathy, Phys. Rev. Lett. 55, 600 (1985)
\bibitem{cmc-15-119}A. V. Ruban and H. L. Skriver Computational Materials Science, 15,
119 (1999)
\bibitem{pr-156-809}P. Soven Phys. Rev. 156, 809 (1967) 
\bibitem{prb-45-13244}J. P. Perdew and Y. Wang, Phys. Rev. B 45, 13244 (1992)
\bibitem{skriver-1984}The LMTO method, Muffin tin orbitals and electronic structure, Hans.
L. Skriver, Springer- Verlag (1984)
\bibitem{prb-38-12807}D. D. Johnson, Phys. Rev. B 38, 12807-12813 (1988)
\bibitem{PRB-56-9319}Abrikosov \emph{et al}., Phys. Rev. B 56, 9319 (1997)
\bibitem{physC-370-211}A. Bharathi \emph{et al.}, Physica C 370, 211 (2002)
\bibitem{physC-387-301}M. Avdeev \emph{et al.},  Physica C 387, 301 (2003)
\end{thebibliography}
\end{document}